# Optimized Implementation of Elliptic Curve Based Additive Homomorphic Encryption for Wireless Sensor Networks

Osman Ugus, Dirk Westhoff
NEC Europe Ltd.
69115 Heidelberg, Germany
{ugus|westhoff}@netlab.nec.de.

Ralf Laue, Abdulhadi Shoufan,
Sorin A. Huss
Integrated Circuits and Systems Lab
Technische Universität Darmstadt, Germany
{laue|shoufan|huss}@iss.tu-darmstadt.de

## ABSTRACT

When deploying wireless sensor networks (WSNs) in public environments it may become necessary to secure their data storage and transmission against possible attacks such as node-compromise and eavesdropping. The nodes feature only small computational and energy resources, thus requiring efficient algorithms. As a solution for this problem the TinyPEDS approach was proposed in [7], which utilizes the Elliptic Curve ElGamal (EC-ElGamal) cryptosystem for additive homomorphic encryption allowing concealed data aggregation. This work presents an optimized implementation of EC-ElGamal on a MicaZ mote, which is a typical sensor node platform with 8-bit processor for WSNs. Compared to the best previous result, our implementation is at least 44% faster for fixed-point multiplication. Because most parts of the algorithm are similar to standard Elliptic Curve algorithms, the results may be reused in other realizations on constrained devices as well.

## 1. INTRODUCTION

Wireless sensor networks (WSNs) are used to monitor certain phenomena in their environment. These phenomena may be temperature in meteorological, vibrations in seismic, or radioactivity in hazardous applications. However, usually the employed devices are only equipped with limited resources in terms of energy, storage, and CPU power. The resource consumption in a WSN is tied to the amount of the data being processed, stored, and transmitted. Therefore, it is necessary to reduce the amount of such data in the network without loosing relevant information.

Depending on the way the data is collected and processed, WSNs are grouped into synchronous and asynchronous. Former ones transmit the sensed data in real-time to a collector device, while latter ones store the sensed data distributed in the WSN and transmit it on demand. This work focuses on asynchronous WSNs only. The device used for requesting and collecting data from the WSN is assumed to be a mobile device such as a laptop, i.e. with relatively high computational resources.

Since WSNs are typically deployed in public and untrusted environments, the data storage and transmissions must be protected, which may be achieved using encryption techniques. In asynchronous WSNs the sensed data needs to be stored in the network until it is collected by the reader device. Therefore, such networks have to provide the functionality of a distributed database as well. The naive way would be to store each data value individually and transmit them when requested. However, this approach does not scale well and collides with the low resource availability of the nodes.

As a solution for these problems, the authors of [7] proposed the tiny persistent encrypted data storage in asynchronous WSNs (TinyPEDS). It incorporates an approach to reduce the data amount by *in-network data aggregation*, i.e. the data is stored in a condensed form. Within the framework of TinyPEDS it is assumed that the use of tamper-resistant hardware for the single nodes of the WSN is prohibitively expensive. Therefore, the application of only symmetric encryption can not guarantee confidentiality, because an attacker may be able to gain the key by examining a node. Because of this, TinyPEDS proposes as a trade-off between security and resource consumption the use of a hybrid approach, namely the combination of a symmetric and a public key encryption, whose sensitive private key is not present in the WSN. The symmetric scheme is used during the data aggregation, while the asymmetric scheme is employed for long-term replicated data storage. However, to facilitate confidential data aggregation, an *additive homomorphic encryption* is needed, for which the Elliptic Curve ElGamal (EC-ElGamal) cryptosystem is used.

This work presents a highly efficient implementation on MicaZ motes for the EC-ElGamal scheme, which is used to realize the public key encryption part of TinyPEDS. Because the WSN nodes have to contain further functionality apart from EC-ElGamal, the ECC implementation should be realized with minimal code size. For the data memory usage a similar motivation holds. In comparison to code and memory size the execution time is not as critical. Therefore, this work focuses on the optimization of code size, memory usage, and computation time – in this order. In comparison with other implementations from literature the proposed solution demands less storage for code, consumes less memory and offers faster operation. Only the solution presented in [9] allows faster execution, however, at the expense of code size. Note that the EC-ElGamal scheme shares many properties with other standard EC algorithms. Thus, the major parts from this work are also applicable to other EC implementations on small general purpose processors.

The following section offers an overview of related work and clarifies the contribution of this work. Section 3 presents a more detailed description of the parts of TinyPEDS relevant for this work. Section 4 discusses the design decisions on the different abstraction levels. The results of the prototype implementation are introduced in Section 5, which also offers some comparisons with other realizations. Section 6, finally, concludes this paper with some closing remarks.

## 2. RELATED WORK

Elliptic Curve implementations for small processors may be found in [9], [11], and [19]. The authors of [9] introduced the utilization of modular arithmetic over $\mathbb{GF}(p)$ with the pseudo-Mersenne prime reduction and the Hybrid multiplication. For further improvements they employed mixed coordinates and the *non-adjacent form* (NAF) reducing the number of needed point additions.

In [11] precomputation using the sliding window method is proposed to speed up the point multiplication, although no signed representation is used. The authors of [19], finally, utilize all above methods including NAF and precomputation of points.

This work is based on these proposals by introducing the Interleave method from [13] and substituting the NAF with the *mutual opposite form* (MOF), see [15]. Employing precomputation of points the Interleave method allows the reduction of needed point doublings, while the MOF exhibits smaller storage requirements. Finally, this work applies the fast pseudo-Mersenne prime reduction not only to the multiplication, but to the addition as well.

## 3. TINYPEDS

TinyPEDS [7] is an approach for asynchronous WSNs, which allows confidential, memory-efficient, and distributed storage of sensed data on resource constrained devices. This work concentrates on the efficient realization of the public key security primitives in TinyPEDS, i.e. the concealed data aggregation using the EC-ElGamal scheme. Thus, for a better understanding of the following sections, this section highlights these parts of TinyPEDS in more detail.

### 3.1 Concealed data aggregation

Because resource consumption – both energy and storage capacity – is critical for the overall lifetime of WSNs, it is necessary to employ techniques to minimize this consumption. In-network data aggregation works toward this goal by condensing the data stored and transmitted, while still providing the necessary amount of information. In TinyPEDS this concentration is done by storing sums of sensed values, which allows the calculation of the average. Figure 1 depicts a simple example for in-network data aggregation, where the sensed values are not encrypted.

However, to limit damage caused by possible attacks the data has to be encrypted. Then, the aggregator node A should calculate the $Sum = [Enc(15) + Enc(16) + Enc(18) + Enc(14)]$ of the encrypted values. This scenario is depicted in Figure 2.

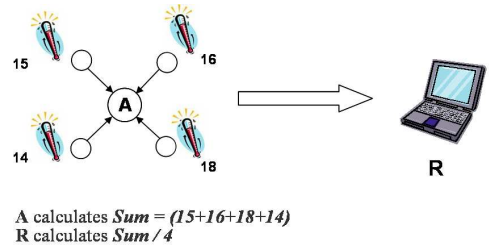

Figure 1: Data aggregation in WSNs

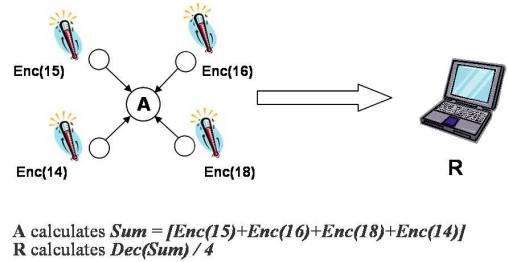

Figure 2: Concealed data aggregation in WSNs

In general, the decrypted sum of ciphertexts $Dec[Enc(15) + Enc(16) + Enc(18) + Enc(14)]$ will not be equal to the sum of the plaintexts $(15 + 16 + 18 + 14)$. Thus, the encryption scheme employed needs to support the following property.

$$Enc(a_1 + a_2 + \ldots) = Enc(a_1) \oplus Enc(a_2) \oplus \ldots,$$

where $Enc(a)$ denotes the encryption of a message $a$ and $\oplus$ represents an addition performed on ciphertexts from a public encryption scheme. An encryption scheme with this property is called *additive homomorphic*.

In [14] several candidates for an asymmetric additive homomorphic encryption schemes were studied. In their analysis, the authors observed that the EC-ElGamal encryption scheme represents the most promising one due to its superior performance and small ciphertext size. Based on this, we decided to use EC-ElGamal within the TinyPEDS framework.

### 3.2 Elliptic Curve ElGamal encryption scheme

The original ElGamal encryption scheme, see [6], is not additive homomorphic. However, the elliptic curve group is an additive group, which can be used to get an additive homomorphic scheme. Algorithm 1 and Algorithm 2 show the methods for EC-ElGamal encryption and decryption, respectively. Therein, $E$ is an elliptic curve over the finite field $\mathbb{GF}(p)$. The order of the curve $E$ is denoted $n = \#E$ and $G$ is the generator point of the curve $E$. The secret key is defined as integer number $x \in \mathbb{GF}(p)$, while the public key is determined as $Y = xG$.

The function $map()$ is a deterministic mapping function used to map values $m_i \in \mathbb{GF}(p)$ to curve points $M_i \in E$ such

**Algorithm 1** EC-ElGamal encryption

**Require:** public key $Y$, plaintext $m$
**Ensure:** ciphertext $(R, S)$
1: choose random $k \in [1, n-1]$
2: $M := map(m)$
3: $R := kG$
4: $S := M + kY$
5: **return** $(R, S)$

**Algorithm 2** EC-ElGamal decryption

**Require:** secret key $x$, ciphertext $(R, S)$
**Ensure:** plaintext $m$
1: $M := -xR + S$
2: $m := rmap(M)$
3: **return** $m$

that

$$map(m_1 + m_2 + \ldots) = \underbrace{map(m_1)}_{M_1} + \underbrace{map(m_2)}_{M_2} + \ldots$$

holds. This is necessary, because the addition over an elliptic curve is only possible with points on that curve, thus, integers have to be mapped to corresponding points. For this purpose, each integer $m$ is mapped to a curve point $M$, where $M$ is the $m$-multiple of the generator point $G$, i.e. $M = mG$. The reverse mapping function $rmap()$ extracts $m$ from a given point $mG$. The mapping function

$$map : m \to mG \text{ with } m \in \mathbb{GF}(p)$$

exhibits the additive homomorphic property, because

$$\begin{aligned} M_1 + M_2 + \ldots &= map(m_1 + m_2 + \ldots) \\ &= (m_1 + m_2 + \ldots)G \\ &= m_1 G + m_2 G + \ldots \end{aligned}$$

holds, where $m_1, m_2, \cdots \in \mathbb{GF}(p)$.

The mapping function is not security relevant, i.e. it neither increases nor decreases the security of the EC-ElGamal encryption scheme. Note that the reverse mapping function is equivalent to solving the *discrete logarithm problem* over an elliptic curve, which represents a computational drawback of this scheme. The reader device must solve $rmap()$ using a brute force approach, because it does not know the sum of the sensed data values $m$ beforehand. However, $rmap()$ is only performed on the powerful reader device and the maximum length of the final aggregation is assumed to be small enough in realistic WSNs (e.g., 3 byte) allowing a successful brute force approach resulting in $m$, see [14].

## 4. DESIGN DECISIONS

Before starting the actual realization of the EC-ElGamal system, several algorithms for its different parts have to be decided on. Thus, the system was divided into four abstraction levels depicted in Figure 3. For each of the lower three levels the employed algorithms are introduced in the following.

### 4.1 Finite field level

This level contains operations like addition or multiplication within the finite field the elliptic curve is defined upon.

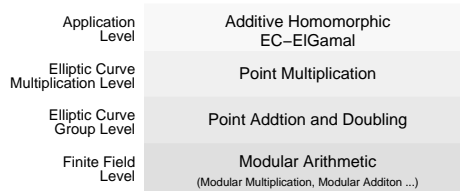

**Figure 3: EC-ElGamal abstraction levels**

Since all operations on higher levels are based on operations on this level, it is critical for performance.

#### 4.1.1 Underlying finite field

Elliptic curve cryptography is usually implemented either over the prime field $\mathbb{GF}(p)$ or the binary field $\mathbb{GF}(2^m)$. In [4] the performance and memory requirements of several elliptic curve algorithms over these fields are studied. The authors found that the number of memory accesses and the code sizes of realizations using binary fields is higher. Furthermore, binary field arithmetic, particularly multiplication, is not well-supported by usual microprocessors, thus, it leads to lower performance, see [9].

However, the inversion in $\mathbb{GF}(p)$ is computationally more expensive than in $\mathbb{GF}(2^m)$. In general, this can somewhat be compensated by using a different coordinate system, in which only one inversion is needed to compute the final result. In the context of TinyPEDS the decryption is only executed on the reader device. Thus, by storing the intermediate results in another coordinate system, the final inversion may be executed on the computationally more powerful reader device. Thus, we decided to use $\mathbb{GF}(p)$ in our implementation.

#### 4.1.2 Multi-precision multiplication

According to [9] 85% of the execution time of a typical point multiplication is spent for multi-precision multiplications on resource constrained devices. Thus, the optimization of the multi-precision multiplication is critical for the overall performance.

Additionally to the well-known *Schoolbook* and *Karatsuba* multiplication, see [12], [18] analyzed the *Comba* multiplication, see [8], and the *Hybrid* multiplication from [9]. The comparison of these approaches presented in [18] shows that the Hybrid method is the most promising one, because it combines the advantages of the Schoolbook and Comba method and needs a small amount of registers and memory accesses.

#### 4.1.3 Modular reduction

After each multi-precision operation a modular reduction is executed to guarantee that the result is in $\mathbb{GF}(p)$. The reduction methods analyzed are the well-known *Montgomery* and *Barrett reduction* and the less general *pseudo-Mersenne prime reduction*, as given in [12]. Although the latter method is far more efficient, this comes at the cost of a special form for possible primes $p$. However, because this does not limit security, it is a good trade-off with respect to the low computational power of the target platform. A good introduction to pseudo-Mersenne prime reduction can be found in [10].

A $n$-bit pseudo-Mersenne prime has the form $p = 2^n - c$, where $n$ is the bit length and $c$ the sum of few powers of 2 with $c \ll 2^n$. The efficiency of the modular reduction stems

from the fact that
$$2^n - c \equiv 0 \mod p \Leftrightarrow 2^n \equiv c \mod p$$

Thus, for the modular reduction, $2^n$ can be substituted with $c$, which is smaller than $2^n$ and therefore results in a residue class with smaller bit-length. Let $a$ and $b$ be $n$-bit integers and $r = a \cdot b$ their $2n$-bit product, which may be represented as

$$r = r_h \cdot 2^n + r_l \equiv r_h \cdot c + r_l \mod p \quad (1)$$

where $r_h$ and $r_l$ denote the $n$ most and least significant bits of $r$, respectively. Since $2^n$ is substituted with the much smaller $c$, $r$ strictly decreases for each substitution, see [12]. The reduction may be computed with only few iterations. The complete algorithm for the reduction after the multiplication may be found in [10].

*4.1.4 Modular addition*

The modular addition $z = a + b$ is executed in two steps. In the first step the not-reduced intermediate result $r = a+b$ is computed. Traditionally, $r$ is reduced in the second step according to

$$z = \begin{cases} r - p & \text{if } r \geq p \\ r & \text{otherwise} \end{cases}$$

By using the pseudo-Mersenne prime reduction instead of the subtraction, the execution time and code size may be decreased, which is of special interest in case of WSNs. Thus, instead of subtracting $p$ the constant $c$ is added, if $r \geq p$. The subtraction of $2^n$ may be done implicitly by ignoring the carry, which either is produced during the addition $a+b$ or the addition $r + c$.

Our investigations showed that the performance of modular addition using the pseudo-Mersenne reduction is nearly twice as good as the usual approach based on subtraction. This is because only the non-zero bytes of the $c$ have to be added. Furthermore, only the bytes, which actually need to be changed, have to be read/written from/to memory.

## 4.2 Elliptic curve group level

This abstraction level provides operations to form an additive group over the elliptic curve points. These operations are point addition, which adds two different points, and point doubling, which adds a point to itself. The result in both cases is again a point on the curve. For a thorough introduction to ECC, see [3].

Elliptic curves may be represented in different coordinate systems, which differ in storage requirements and computational effort for the algorithms of point addition and doubling. A good overview on different coordinate systems is offered in [5], which was used for the investigation in [18].

*Affine*, *Projective*, *Jacobian*, *Chudnovsky-Jacobian*, and the *Modified Jacobian* coordinate systems were compared in terms of performance and memory requirements. As a further possibility [5] suggests the use of *mixed* coordinate systems, i.e. input and output points are represented in different coordinate systems. The analysis in [18] showed that this idea allows better results than all above coordinate systems considered separately. However, because TinyPEDS exhibits different constraints, e.g. no finite field inversion is needed, the examination in [18] leads to different results than in [5].

According to [18] $\mathcal{AJJ}$ coordinates for point addition exhibit the smallest storage requirements and computational effort. In this notation, $\mathcal{A}$ and $\mathcal{J}$ denote the affine and Jacobian coordinate system, respectively. Furthermore, $\mathcal{AJJ}$ means for a point addition that the first input point is given in affine coordinates, the second input point is given in Jacobian coordinates, and the output point is computed in Jacobian coordinates. For the point doubling, in turn, the form $\mathcal{JJ}$ is suggested, which means that both input and output points are represented in Jacobian coordinates.

## 4.3 Elliptic curve multiplication level

The main operation in elliptic curve cryptosystems is the scalar or point multiplication. Although its efficiency depends on the operations on the lower levels, the algorithm on this level still has a high impact on overall execution time. The optimization of the basic *Double and Add* algorithm depends mainly on its direction. For further optimizations the number of doublings may be reduced with the *Interleave method*, while *signed representations*, in contrast, allow the reduction of the number of point additions. Note that the Interleave method and the general signed representations, i.e. wMOF with $w > 2$, require precomputation of points, which is only feasible for scalar multiplications with a fixed base point $G$. If the base point changes for every point multiplication, precomputation results in a higher computational overhead.

The double and add algorithm is the elliptic curve complement to the square and multiply algorithm used for exponentiation, see [12]. Both *Left-to-Right* and *Right-to-Left* directions are possible. According to [18] the Left-to-Right method is superior due to lower storage requirements and the fact, that the Right-to-Left method does not allow $\mathcal{AJJ}$ point additions.

*4.3.1 Reducing the number of point doublings*

The Interleave method was originally proposed for multi-scalar multiplications of the form $(k_1 \cdot P_1 + k_2 \cdot P_2 + \ldots)$, see [13]. However, this idea can also be used to perform single scalar multiplications. This is due to the fact that the scalar multiplication $kP$ of an $n$-bit scalar $k$ with a curve point $P$ may also be represented as a sum of $t$ partial multiplications with $t$ scalars $k_i$ and $t$ points $P_i$ as follows.

$$kP = (k_t \cdot 2^{(t-1)n/t} + ... + k_2 \cdot 2^{n/t} + k_1) \cdot P \quad (2)$$
$$= \sum_{i=1}^{t} k_i \cdot P_i \quad (3)$$

Thereby, each scalar $k_i$ is a $n/t$-bit long part of $k$ and $P_i = 2^{(i-1)n/t}P$. Note that $n$ is padded with 0's from left until it is a multiple of $t$.

The Interleave method leads to a performance increase, if the points $P_i$ are precomputed and stored off-line, as shown in Algorithm 3. Therein, $ECADD$ and $ECDBL$ refer to the point addition and doubling, respectively.

*4.3.2 Reducing the number of point additions*

Because a point addition is only executed if the corresponding bit of the scalar $k$ is 1, a signed representation which reduces the *Hamming weight* of $k$ can be used to reduce the number of point additions. [18] analyzed both the *non-adjacent form* (NAF) and the *mutual opposite form* (MOF), see [16] and [15], respectively. Although both lead to the same Hamming weight, MOF is superior to NAF, because it exhibits lower memory requirements.

**Algorithm 3** Interleave method with precomputed points

---
**Require:** $n$-bit scalar $k$ split into $t$ parts $k_i$ with $\frac{n}{t}$ bit each, precomputed points $P_i = 2^{(i-1)n/t}P$ for $i \in [2, t]$, $n$ is a multiple of $t$

**Ensure:** scalar product $R = kP$
1: $R := \emptyset$
2: **for** $j$ from $(\frac{n}{t} - 1)$ to $0$ **do**
3:    $R := ECDBL(R)$
4:    **for** $i$ from $1$ to $t$ **do**
5:       **if** $k_{i_j} == 1$ **then** {$k_{i_j}$ denotes the $j$th bit of $k_i$}
6:          $R := ECADD(R, P_i)$
7:       **end if**
8:    **end for**
9: **end for**
10: **return** $R$

---

The Hamming weight of $k$ represented in $w$MOF is $1/(1+w)$ on average, where $w$ is the bit width of the used bytes, see [15]. Therefore, the amount of needed point additions is reduced by $1+w$. Moreover, because negative points $-Q = -(Q) = -(x, y) = (x, -y)$ do not have to be precomputed, the amount of precomputed points for the $w$MOF method is $2^{w-2} - 1$. This leads to a total of

$$(t - 1) + t \cdot (2^{w-2} - 1)$$

precomputed points, if both Interleave method and $w$MOF are used.

## 5. PROTOTYPE IMPLEMENTATION

For the prototype implementation a MicaZ mote, see [1], was used, which is a typical device for WSNs and is equipped with a 8-bit processor. The bit width of coordinates and scalar $k$ was chosen to be 160, thus the multi-precision integers need 20 bytes each. The realization employed TinyOS-2.0, an open-source operating system designed for WSNs, see [2]. The prototype was implemented with *nesC* V1.2.8a and compiled using *gcc* V3.4.3 for the AVR processor using the default compiler options specified in TinyOS-2.0 make system. The operations on the lowest abstraction level were realized using Assembler, while those on higher levels were written in nesC. The timing results form the prototype are the average over 500 executions with random numbers and were generated using the curve parameters *secp160r1* from [17].

The performance of different point multiplications is shown in Table 1, which also presents performance values for solutions from literature. The upper part contains multiplications without precomputed points, while the lower part those with precomputation. The point multiplication in the first row was executed with the simple Left-to-Right binary method, while 2MOF was employed for the second row. The point multiplications in the lower part of Table 1 from this work employed the Interleave method.

Note that the performance values for TinyECC-0.2 in Table 1 were obtained by integrating it into the test suite from this work, because the values were not presented in [11]. Furthermore, this work does not contain the conversion back to affine coordinates including the expensive modular inversion, because it is executed on the reader device in TinyPEDS. According to [5], the ratio between modular multiplications and inversions $Inv/Mult$ is 30. Thus, together with the additional four modular multiplications needed for the conversion, this results in $34 \cdot 0.532ms = 18ms$, where $0.532ms$ is the execution time of one modular multiplication, see [18]. Therefore, for a fair comparison, this value is already added to the results in Table 1.

**Table 1: Comparison of point multiplications**

| Reference | #Prec. points | Exec. time | Code size [bytes] | Memory size [bytes] |
|---|---|---|---|---|
| This work | 0 | 1.23s | 2096 | 180 |
| This work (2MOF) | 0 | 1.03s | 2794 | 260 |
| [9] | 0 | 0.81s | 3682 | 282 |
| [19] | 0 | 1.35s | n/a | n/a |
| [11] | 0 | 1.78s | 6562 | 382 |
| This work | 1 | 0.69s | 3166 | 481 |
| This work | 2 | 0.57s | 3536 | 543 |
| [19] | 15 | 1.24s | n/a | n/a |

**Table 2: Implementation results for EC-ElGamal**

| #Precomputed points | Execution time | Code size [bytes] | Memory size [bytes] |
|---|---|---|---|
| 0 | 2.48s | 2726 | 320 |
| 0 (2MOF) | 2.16s | 3172 | 400 |
| 2 | 1.42s | 3806 | 621 |
| 4 | 1.19s | 5122 | 683 |

Table 2, finally, shows the performance of different realizations of the EC-ElGamal, which contains each two point multiplications with an $n$-bit scalar $k$ and one short point multiplication with the sensed data $m$, see Algorithm 1. Note that for test purposes $m$ is chosen to be 8-bit.

### 5.1 Assessment of non-precomputation

For the comparison with the point multiplication from [9], we used our solution from the second row of Table 1, because both use a signed representation. Our solution is about 21% slower, but its code size is about 24% smaller. For this we have several possible explanations. Firstly, we applied *loop unrolling* only to loops most critical for execution time in order to keep the code size small. However, we believe that the authors of [9] employed loop unrolling for every loop in the finite field and elliptic curve operations. Secondly, to reduce code-size, we executed modular squaring with multiplications, rather than using a dedicated squaring operation. Thirdly, the solution from [9] was completely implemented in Assembler, while we did this only for the finite field operations and wrote the elliptic curve operations in nesC. Therefore, we believe that the performance of our elliptic curve operations may be further improved by implementing them in Assembler. Finally, our realization uses TinyOS, which probably consumes additional CPU cycles.

Compared with the solution from [11], which is also executed on a MicaZ mote, our implementation is 42% faster and its code size is 57% smaller. The performance gains of our solution presumably stem from more optimized modular operations and the use of the 2MOF representation enabling a faster point multiplication.

Compared with the solution from [19], finally, our implementation performs 23% faster. The authors did not present figures about code and memory usage of their solution. However, because their implementation employed similar acceleration techniques as used in TinyECC-0.2 and the NAF representation, we believe that our solution is superior to [19] in terms of code size and memory consumption.

## 5.2 Assessment of precomputation

Although the implementation from [19] uses 15 precomputed points, it is significantly slower than both our solutions employing only 1 and 2 precomputed points (see Table 1). Our solution is 44% and 54% faster, when 1 and 2 precomputed points are used, respectively. Again, because [19] employs similar acceleration techniques as TinyECC-0.2, we believe that our realization is superior in terms of code size as well.

In the design from [19] the number of the precomputed points must be $2^s - 1$, where $s$ is the window size. This limits possible numbers of precomputed points, while the Interleave method from our realization allows an arbitrary amount with 2MOF representation. Furthermore, in our case the Interleave method yields a higher benefit per precomputed point, because it divides the number of point doublings by $t$, while the precomputation method from [19] reduces the – usually smaller – number of point additions to $(1/s) \cdot (1 - 1/2^s)$. Therefore, our solution gains a higher benefit from the same memory usage increase than the solution from [19].

## 6. CONCLUSION

We presented an optimized realization of an elliptic curve based additive homomorphic encryption to be used in the TinyPEDS framework in WSNs. The main optimization techniques employed are pseudo-Mersenne prime reduction on the finite field level and the Interleave method and the signed MOF representation for the point multiplication. Our implementation offers a fast point multiplication, while featuring small code and memory requirements. Finally, although it was designed for the TinyPEDS framework, it may be also used for more general elliptic curve implementations.

## 7. ACKNOWLEDGMENTS

The work presented in this paper is supported in part by the European Commission within the STREP UbiSec&Sens. The views and conclusions contained herein are those of the authors and should not be interpreted as necessarily representing the official policies or endorsements, either expressed or implied, of the UbiSec&Sens project or the European Commission. This work was also partly supported by the research program SicAri of BMBF (German Federal Ministry of Education and Research).